# Stuffed Rare Earth Pyrochlore Solid Solutions

G. C. Lau, B. D. Muegge, T. M. McQueen, E. L. Duncan, and R. J. Cava

*Department of Chemistry, Princeton University, Princeton NJ 08544*

## Abstract

Synthesis and crystal structures are described for the compounds $Ln_2(Ti_{2-x}Ln_x)O_{7-x/2}$, where *Ln* = Tb, Dy, Ho, Er, Tm, Yb, Lu, and *x* ranges from 0 to 0.67. Rietveld refinements on X-ray powder diffraction data indicate that in Tb and Dy titanate pyrochlores, extra $Ln^{3+}$ cations mix mainly on the $Ti^{4+}$ site with little disorder on the original $Ln^{3+}$ site. For the smaller rare earths (Ho-Lu), stuffing additional lanthanide ions results in a pyrochlore to defect fluorite transition, where the $Ln^{3+}$ and $Ti^{4+}$ ions become completely randomized at the maximum (x=0.67). In all of these *Ln*-Ti-O pyrochlores, the addition of magnetic $Ln^{3+}$ in place of non-magnetic $Ti^{4+}$ adds edge sharing tetrahedral spin interactions to a normally corner sharing tetrahedral network of spins. The increase in spin connectivity in this family of solid solutions represents a new avenue for investigating geometrical magnetic frustration in the rare earth titanate pyrochlores.

## Introduction

Pyrochlore materials, with general formula $A_2B_2O_7$, represent an important structure type in view of geometrical magnetic frustration due to the ordering of the *A* and *B* cations into separate interpenetrating lattices of corner sharing tetrahedra. Spins of magnetic atoms placed in either lattice position can result in frustration, where the presence of many energetically equivalent ground states suppresses long range ordering of the moments.[1] Unusual low temperature physics as a consequence of magnetic frustration has been seen in the well studied rare earth pyrochlores including cooperative paramagnetism in $Tb_2Ti_2O_7$[2-5] and spin ice characteristics in $Dy_2Ti_2O_7$[6-10] and $Ho_2Ti_2O_7$.[11-16] Novel magnetic states as a result of frustration are also seen in $Y_2Ru_2O_7$[17, 18] and superconducting $RbOs_2O_6$.[19] Most studies focusing on magnetic frustration in pyrochlores are based on magnetic moments isolated on either the *A* or *B*-site, with spin interactions becoming more complicated when both cations are magnetic.[20, 21] While the effect of decreasing the magnetic lattice connectivity through diluting the magnetic atom with non-magnetic species on the same site has been explored,[5, 22, 23] there have not been, to our knowledge, examples of increasing this connectivity by "stuffing" more of the same magnetic atom onto the non-magnetic site.

Stuffing more *A* cations onto the *B*-site in pyrochlores can lead to disordering of the separate corner-sharing tetrahedral networks. This disordering, or cation mixing, typically occurs only in pyrochlores with larger *B* cations, such as $Zr^{4+}$.[24-27] For the rare earth titanate pyrochlores, due to the large difference in size between $Ln^{3+}$ and $Ti^{4+}$ ions, mixing of cations in the $Ln_2Ti_2O_7$ stoichiometry has only been reported under conditions such as ball milling, ion irradiation, or reduction with $CaH_2$.[28-30] Stuffed cubic phases of titanates, in the form $Ln_2(Ti_{2-x}Ln_x)O_{7-x/2}$ are reported for $Ln$ = Dy-Lu, but only at limited stuffing levels with the focus on these materials as oxygen ionic conductors.[31-37] According to the early phase diagrams, a wider range of $Ln^{3+}$ stuffing is possible for cubic titanates of the smaller lanthanides (Ho-Lu) at high temperatures, but these stoichiometries are difficult to obtain as different crystal structures are more thermodynamically stable at low temperature.[38-44]

Here we report the synthesis and crystal structures determined by X-ray powder diffraction of polycrystalline $Ln_2(Ti_{2-x}Ln_x)O_{7-x/2}$, where $Ln$ = Tb-Lu and x ranges from 0 to 0.67. These solid solutions in the $Ln_2O_3$-$TiO_2$ systems contain $Ln$/Ti ratios ranging from 1, in the original pyrochlore, to 2 in the fully stuffed materials. The occupation of magnetic $Ln^{3+}$ ions in either $A$ or $B$ site is determined with respect to the stuffing level, illustrating the structural evolution in these compounds as the connectivity in the magnetic spins is increased. The placing of additional spins in the corner sharing tetrahedral network potentially opens a new window in the study of rare earth titanates in light of geometrical magnetic frustration.[45]

## Experimental

To synthesize $Ln_2(Ti_{2-x}Ln_x)O_{7-x/2}$ ($Ln$ = Tb-Lu), stoichiometric amounts of rare earth oxides ($Tb_4O_7$, Johnson Matthey, 99.9%; $Dy_2O_3$, Alfa Aesar, 99.9%; $Ho_2O_3$, Alfa Aesar, 99.9%; $Er_2O_3$, Alfa Aesar, 99.99%; $Tm_2O_3$, Alfa Aesar, 99.9%; $Yb_2O_3$, Aldrich, 99.9%; $Lu_2O_3$, Alfa Aesar, 99.99%) and $TiO_2$ (Aldrich,99.9%) were intimately ground and pressed into pellets. For $Ln$ = Ho-Lu, samples were heated in a static argon atmosphere at 1700 °C for 12 hours, and quenched in a water cooled furnace chamber to room temperature in approximately 30 minutes. Stuffed titanates with Tb and Dy were made by arc-melting 200 mg sized pellets to allow for faster quenching of the samples.

The crystal structures were characterized through X-ray powder diffraction (XRD) using CuKα radiation and a diffracted beam monochromator. The software TOPAS 2.1 (Bruker AXS) operated in the fundamental parameters approach was used for Rietveld structure refinements. All patterns were refined in space group Fd-3m (No. 227) with the origin placed on the 16c site (*B*-site). $Ln^{3+}$ and $Ti^{4+}$ occupancies were allowed to refine freely on both the *A* and *B* sites of the pyrochlore, with the constraint that their total occupancies from both sites maintained the nominal stoichiometric ratios. Oxygen occupation on both the 8b and 8a sites was allowed to refine with the constraint that the total occupancy on both sites summed to maintain charge neutrality in the compound as there was no indication of lower oxidation state Ti in any of the phases. The oxygen thermal parameters were fixed at reasonable values, given the well known insensitivity of the XRD powder refinement method to the details of the oxygen array in heavy ion compounds. In addition to the parameters shown in Table 1, other refined parameters include background, zero error correction, scaling factor, and crystal size and strain. For the case of the arc melted samples, which had well developed crystallites, a preferred orientation parameter was included in the refinements.

Electron diffraction experiments were performed at room temperature with a Philips CM200ST electron microscope equipped with a 200 kV field emission gun. Electron diffraction was performed using a condenser aperture of 10 μm and an electron probe size of 10 nm diameter. Electron transparent samples were obtained by crushing the specimens under acetone, and depositing a few droplets on a holey carbon Cu grid.

## Results and Discussion

In general, the thermodynamic phase diagrams[38-42, 46-48] between the compositions $Ln_2Ti_2O_7$ and $Ln_2TiO_5$ show full two phase regions between the fixed composition end members, with hexagonal and orthorhombic stable ordered phases at low temperatures in ordered $Ln_2TiO_5$. Disordered high temperature fluorite type $Ln_2TiO_5$[38-42, 46] and limited solid solutions[44] are reported for Dy-Lu lanthanides. In the current study, however, the solid solution series, $Ln_2(Ti_{2-x}Ln_x)O_{7-x/2}$ ($Ln$ = Tb-Lu, $0 \leq x \leq 0.67$), all refined with the cubic pyrochlore structure, were successfully synthesized as single phase materials through high temperature synthesis followed by rapid quenching. Quenching allowed for the bypassing of more thermodynamically stable low temperature phases and phase separation into rare earth richer and poorer phases. $Ln_2(Ti_{2-x}Ln_x)O_{7-x/2}$ solid solutions based on the smaller lanthanides (Ho-Lu), with $Ln^{3+}$ closer in size to $Ti^{4+}$, formed with a slower quench (30 minutes to room temperature from 1700 °C) whereas stuffed pyrochlore titanates based on Tb and Dy, the largest successful cases, required arc melting to cool more quickly. The x=0.4 and 0.5 pyrochlores in the Tb series were never successfully formed as pure single phases, having the more stable, ordered hexagonal symmetry phase also present. In the Gd-Ti-O system and for larger lanthanides, even the fast cooling rates attained in arc melting could not stabilize any substantial stuffing level in the cubic pyrochlore structure. This difficulty in forming the stuffed solid solutions for the larger rare earths is consistent with the expectation that ions with large size differences (i.e. large $Ln^{4+}$ and $Ti^{4+}$) will tend not to mix on the same site. Simplified phase diagrams comparing the high temperature phases known in the literature[38-42, 46-48] with those found in this study for each rare earth titanate are presented in Figure 1.

The final refined structural parameters for all the stuffed pyrochlores studied are presented in Table 1. The refinement agreement parameters indicate that the structural models are excellent representations of the structures. For the smaller rare earths, Ho-Lu, a smooth pyrochlore to defect fluorite transition is seen as a function of stuffing. As shown in Figure 2, with $Ho_2(Ti_{2-x}Ho_x)O_{7-x/2}$ as an example, the x=0 non-stuffed pyrochlore begins with ordered corner sharing tetrahedral lattices of both $Ho^{3+}$ and $Ti^{4+}$. The magnetic $Ho^{3+}$ atoms on the *A*-site form a distinct network separated from the non-magnetic $Ti^{4+}$ on the *B*-site by a crystallographic offset of (½, ½, ½), yet the two sublattices are exactly equivalent in size and atomic distances. As more $Ho^{3+}$ is added in the range $0 \leq x \leq 0.3$, $Ho^{3+}$ replaces $Ti^{4+}$ strictly on the *B*-site (Table 1), developing edge-sharing tetrahedral spin interactions outlined in green. Above x=0.4, $Ti^{4+}$ begins to mix onto the originally pure $Ho^{3+}$ 16d site. At x=0.67, the maximally doped sample has complete mixing of $Ho^{3+}$ and $Ti^{4+}$ so that the *A* and *B*-sites are indistinguishable from each other. The lack of cation ordering reduces the pyrochlore superstructure to the smaller unit celled sized fluorite structure. The oxygen atoms are not considered in detail in the present work despite their important role in stabilizing the pyrochlore to fluorite structure transition[24] due to the insensitivity of our powder XRD method to that information. Future studies by neutron diffraction designed to determine the manner in which the oxygen ion distribution changes during the pyrochlore to fluorite transition in these phases will be of considerable interest.

Figure 3a illustrates the linear increase of the cubic pyrochlore lattice parameter, *a*, with excess lanthanide concentration, indicative of the successful stuffing of extra $Ln^{3+}$ on the B site in each series. The inset shows that the rates of increase in the unit cells (i.e. the slopes of *a* versus *x*) also change linearly with radius of the different rare earth cations (effective radii of $Ln^{3+}$ ions in six-fold oxygen coordination as given by Shannon and Prewitt).[49] At given compositions, *x*, the lattice parameters continue to generally follow a linear increase with lanthanide radius as shown in Figure 3b.

Figure 4 plots the $Ti^{4+}$ occupation of the originally pure $Ln^{3+}$ *A*-site as a function of $Ln^{3+}$ doping, x. In the range $0 \leq x \leq 0.3$, all the lanthanides (Tb-Lu) show virtually no mixing within the experimental uncertainties associated with our XRD method. Above x=0.4, the smaller Ho-Lu series show $Ti^{4+}$ mixing continuously and rapidly onto the *A*-site to about a 30% level – the expected disordered ratio in the fluorite structure with chemical composition $Ln_2TiO_5$. For Tb and Dy, the *A*-site mixing increases slightly (< 10%) but never reaches the fully disordered level required in fluorite and thus maintains a pyrochlore-like structure through x=0.67 stuffing. Figure 5 shows an example of one of the observed and calculated powder XRD patterns for one of the small lanthanide (Ho) stuffed pyrochlores, and the inset shows a detail of one of the diffraction peaks across the series, indicating the presence of good crystallinity.

Difficulties in synthesizing Tb and Dy stuffed pyrochlores are represented in Figure 6. The main panel shows a structure refinement of the x=0.1 stuffed Dy titanate with sharp symmetric peaks of a well crystallized polycrystalline material. However, upon further doping of extra Dy, the inset shows how the diffraction peaks become broader and asymmetric. The Tb patterns display the same characteristics. We interpret the broadening of the peaks to be a sign of compositional inhomogeneity in the samples caused by the thermodynamically driven phase separation at low temperatures to pyrochlore $Dy_2Ti_2O_7$ and fluorite-like $Dy_2TiO_5$. An estimate of the spread in composition can be made from the excess peak width and the variation of the peak position with average composition. The spread is determined to be x=0.58 to 0.67 for the maximally stuffed Tb sample and x=0.62 to 0.67 for Dy. Difficulties in fitting the broadened peaks in this family did not significantly impact the determination of the metal atom mixing, which is the main structural characterization undertaken in the present study.

The very large difference in sizes between $Ln^{3+}$ and $Ti^{4+}$, even for the small lanthanides, makes the extensive occurrence of the stuffed pyrochlores quite surprising. To test for the possible presence of short range *Ln*-Ti ordering, electron diffraction was used for the largest lanthanide series that could be fully stuffed, Ln=Dy. Figure 7 gives electron diffraction patterns of x=0.2 and x=0.67 samples in the stuffed Dy-Ti-O series. The sharp ordered spots and absence of diffuse scattering indicate that there are no large regions of short range Dy-Ti ordering occurring in this family.

## Conclusions

We have successfully synthesized polycrystalline samples in a range of stuffed rare earth titanates, $Ln_2(Ti_{2-x}Ln_x)O_{7-x/2}$ ($Ln$ = Tb-Lu, $0 \le x \le 0.67$) by high temperature synthesis and rapid cooling and analyzed their crystal structures. For the smaller lanthanides (Ho-Lu) a smooth pyrochlore to defect fluorite transition is observed as more $Ln^{3+}$ ions are stuffed into the structure. Tb and Dy titanates are more difficult to stuff, given the larger size differences between the $Ln^{3+}$ and $Ti^{4+}$ ions. This results in less mixing between the two cations, with disordering mostly on the *B*-site of the pyrochlore structure, but no transition to fluorite even at maximum stuffing. In all cases, the spin interactions limited to corner sharing tetrahedra in the original pyrochlores are transformed to an edge sharing tetrahedral geometry upon stuffing. The ability to chemically alter the magnetic connectivity in these well known and studied materials presents an opportunity to uncover more unusual behavior associated with geometrical magnetic frustration in future physical characterization.

**Table 1.** Crystallographic parameters for the series $Ln_2(Ti_{2-x}Ln_x)O_{7-x/2}$ in space group *Fd-3m* (No 227). $Ln^{3+}$ and $Ti^{4+}$ both refined on 16d (1/2, 1/2, 1/2) and 16c (0, 0, 0) sites. $O^{2-}$ placed on 8b (3/8, 3/8, 3/8), 8a (1/8, 1/8, 1/8), and 48f (*x*, 1/8, 1/8) sites.

| ***Ln*/ Parameter** | | x: 0 | 0.1 | 0.2 | 0.3 | 0.4 | 0.5 | 0.67 |
|---|---|---|---|---|---|---|---|---|
| **Tb** | | | | | | | | |
| | a (Å) | 10.1694(2) | 10.193(8) | 10.2317(2) | 10.2575(2) | - | - | 10.4022(8) |
| x pos. | 48f | 0.3287(7) | 0.3234(8) | 0.3290(8) | 0.3318(9) | - | - | 0.349(2) |
| $Ti^{4+}$ Occ. | 16d | 0.035(3) | 0.024(3) | 0.035(4) | 0.029(4) | - | - | 0.065(8) |
| | 16c | 0.965(3) | 0.926(3) | 0.865(4) | 0.821(4) | - | - | 0.605(8) |
| $O^{2-}$ Occ. | 8b | 1 | 0.97(2) | 0.87(2) | 0.78(2) | - | - | 0.49(3) |
| | 8a | 0 | -0.02(2) | 0.03(2) | 0.07(2) | - | - | 0.18(3) |
| $B_{eq}$ | 16d | 0.46(5) | -0.01(5) | 0.94(7) | 0.84(7) | - | - | 1.1(2) |
| | 16c | 0.9(1) | 0.9(1) | 1.9(1) | 1.6(1) | - | - | 1.4(2) |
| | $R_{wp}$ | 12.01 | 12.25 | 12.38 | 11.78 | - | - | 13.56 |
| | $\chi^2$ | 1.11 | 1.07 | 1.14 | 1.12 | - | - | 1.23 |
| **Dy** | | | | | | | | |
| | a (Å) | 10.136(1) | 10.1622(2) | 10.1946(2) | 10.2294(4) | 10.2715(4) | 10.3080(5) | 10.3614(6) |
| x pos. | 48f | 0.3194(8) | 0.3265(8) | 0.3319(8) | 0.335(1) | 0.341(1) | 0.331(1) | 0.349(2) |
| $Ti^{4+}$ Occ. | 16d | 0.036(3) | 0.019(3) | 0.018(3) | 0.041(4) | 0.042(5) | 0.042(5) | 0.101(8) |
| | 16c | 0.964(3) | 0.931(3) | 0.882(3) | 0.809(4) | 0.758(5) | 0.708(5) | 0.569(8) |
| $O^{2-}$ Occ. | 8b | 1 | 0.95(1) | 0.90(2) | 0.84(2) | 0.73(2) | 0.65(2) | 0.63(4) |
| | 8a | 0 | 0.00(1) | 0.00(2) | 0.01(2) | 0.07(2) | 0.10(2) | 0.04(4) |
| $B_{eq}$ | 16d | 0.26(5) | 0.18(4) | 0.59(5) | -0.08(7) | 1.1(1) | 0.5(1) | 0.4(2) |
| | 16c | 0.8(1) | 0.9(1) | 1.2(1) | 1.1(1) | 1.4(2) | 0.1(2) | 0.4(2) |
| | $R_{wp}$ | 11.62 | 10.78 | 10.91 | 12.10 | 12.31 | 12.42 | 12.42 |
| | $\chi^2$ | 1.15 | 1.09 | 1.12 | 1.13 | 1.13 | 1.12 | 1.13 |
| **Ho** | | | | | | | | |
| | a (Å) | 10.1048(2) | 10.133(2) | 10.163(2) | 10.1958(5) | 10.2302(4) | 10.2575(3) | 10.3058(3) |
| x pos. | 48f | 0.3294(5) | 0.3271(5) | 0.3264(6) | 0.3336(6) | 0.3361(6) | 0.3378(7) | 0.3367(8) |
| $Ti^{4+}$ Occ. | 16d | 0.016(3) | 0.014(3) | -0.030(3) | -0.002(4) | 0.032(4) | 0.075(5) | 0.351(8) |
| | 16c | 0.984(3) | 0.936(3) | 0.930(3) | 0.852(4) | 0.768(4) | 0.67(5) | 0.320(8) |
| $O^{2-}$ Occ. | 8b | 1 | 0.937(1) | 0.90(1) | 0.79(1) | 0.74(1) | 0.52(2) | 0.70(2) |
| | 8a | 0 | 0.013(9) | 0.00(1) | 0.06(1) | 0.06(1) | 0.23(2) | -0.03(2) |
| $B_{eq}$ | 16d | 0.83(4) | 1.17(4) | 0.63(9) | 1.31(7) | 2.05(8) | 4.2(1) | 0.8(2) |
| | 16c | 1.37(9) | 1.61(8) | 0.41(9) | 1.0(1) | 0.9(1) | -0.7(1) | 1.2(2) |
| | $R_{wp}$ | 9.81 | 8.82 | 9.62 | 9.69 | 9.69 | 10.31 | 10.53 |
| | $\chi^2$ | 1.18 | 1.10 | 1.12 | 1.11 | 1.21 | 1.29 | 1.26 |
| **Er** | | | | | | | | |
| | a (Å) | 10.076(3) | 10.116(6) | 10.132(9) | 10.164(3) | 10.2002(3) | 10.2222(4) | 10.2663(3) |
| x pos. | 48f | 0.3283(5) | 0.3296(6) | 0.3311(5) | 0.3321(7) | 0.3344(6) | 0.347(1) | 0.355(2) |
| $Ti^{4+}$ Occ. | 16d | 0.031(2) | 0.009(3) | 0.004(2) | 0.045(3) | 0.029(3) | 0.146(5) | 0.32(3) |
| | 16c | 0.969(2) | 0.941(3) | 0.896(2) | 0.805(3) | 0.771(3) | 0.604(5) | 0.35(2) |
| $O^{2-}$ Occ. | 8b | 1 | 0.95(1) | 0.89(1) | 0.68(1) | 0.68(1) | 0.51(2) | 0.24(9) |
| | 8a | 0 | 0.00(1) | 0.01(1) | 0.17(1) | 0.12(1) | 0.24(2) | 0.43(9) |
| $B_{eq}$ | 16d | 1.07(4) | 2.26(5) | 1.72(5) | 1.79(7) | 3.42(9) | 0.8(1) | 0.5(6) |
| | 16c | 2.2(1) | 2.3(1) | 1.8(1) | 2.9(1) | 1.7(1) | 0.8(1) | 0.2(6) |
| | $R_{wp}$ | 7.99 | 8.04 | 7.73 | 7.91 | 7.75 | 7.94 | 7.96 |
| | $\chi^2$ | 1.21 | 1.18 | 1.22 | 1.17 | 1.22 | 1.19 | 1.21 |
| **Tm** | | | | | | | | |
| | a (Å) | 10.054(1) | 10.0843(4) | 10.1104(4) | 10.1338(4) | 10.1595(2) | 10.1846(5) | 10.2245(5) |
| x pos. | 48f | 0.3261(5) | 0.3300(7) | 0.3291(7) | 0.3318(7) | 0.3331(8) | 0.347(1) | 0.338(1) |
| $Ti^{4+}$ Occ. | 16d | 0.002(2) | -0.029(3) | -0.010(3) | -0.010(3) | 0.026(3) | 0.137(5) | 0.322(9) |
| | 16c | 1.002(2) | 0.979(3) | 0.910(3) | 0.860(3) | 0.774(3) | 0.613(5) | 0.349(9) |
| $O^{2-}$ Occ. | 8b | 1 | 0.89(1) | 0.83(1) | 0.77(1) | 0.72(2) | 0.55(2) | 0.64(5) |
| | 8a | 0 | 0.06(1) | 0.07(1) | 0.08(1) | 0.08(2) | 0.20(2) | 0.03(5) |
| $B_{eq}$ | 16d | 0.23(3) | 0.16(3) | 0.48(4) | 0.65(4) | 0.88(5) | 1.2(1) | 1.6(2) |
| | 16c | 0.72(8) | 0.21(9) | 0.55(9) | 0.44(9) | 0.35(9) | 0.1(1) | 0.8(2) |
| | $R_{wp}$ | 8.85 | 9.44 | 9.24 | 9.28 | 9.25 | 10.65 | 9.56 |
| | $\chi^2$ | 1.27 | 1.22 | 1.18 | 1.24 | 1.20 | 1.45 | 1.27 |
| **Yb** | | | | | | | | |
| | a (Å) | 10.0356(2) | 10.056(1) | 10.082(2) | 10.102(3) | 10.127(2) | 10.1557(2) | 10.1881(2) |
| x pos. | 48f | 0.3231(5) | 0.3259(6) | 0.3321(7) | 0.3311(7) | 0.3426(8) | 0.346(1) | 0.341(1) |
| $Ti^{4+}$ | 16d | 0.032(2) | -0.026(2) | -0.008(3) | -0.021(3) | 0.078(3) | 0.150(6) | 0.329(8) |
| Occ. | 16c | 1.032(2) | 0.976(2) | 0.908(3) | 0.871(3) | 0.722(3) | 0.520(6) | 0.341(8) |
| $O^{2-}$ | 8b | 1 | 0.95(1) | 0.85(1) | 0.82(1) | 0.69(2) | 0.43(3) | 0.67(4) |
| Occ. | 8a | 0 | 0.00(1) | 0.05(1) | 0.03(1) | 0.11(2) | 0.32(3) | 0.00(4) |
| $B_{eq}$ | 16d | -0.09(2) | 1.32(3) | 1.52(4) | 1.07(4) | 1.61(6) | 2.4(2) | 1.5(2) |
| | 16c | 0.41(7) | 1.06(9) | 1.3(1) | 0.42(9) | 1.02(9) | -0.8(1) | 1.2(2) |
| | $R_{wp}$ | 12.87 | 10.01 | 10.05 | 11.72 | 12.72 | 14.87 | 11.28 |
| | $\chi^2$ | 1.32 | 1.17 | 1.15 | 1.28 | 1.44 | 1.61 | 1.28 |

**Lu**

| | | | | | | | | |
|---|---|---|---|---|---|---|---|---|
| | a (Å) | 10.0203(4) | - | - | 10.0864(3) | - | - | 10.1683(4) |
| x pos. | 48f | 0.3226(7) | - | - | 0.3335(8) | - | - | 0.340(1) |
| $Ti^{4+}$ | 16d | 0.00(1) | - | - | 0.035(3) | - | - | 0.324(8) |
| Occ. | 16c | 1.00(1) | - | - | 0.815(3) | - | - | 0.346(8) |
| $O^{2-}$ | 8b | 1 | - | - | 0.77(2) | - | - | 0.67(4) |
| Occ. | 8a | 0 | - | - | 0.08(2) | - | - | 0.00(4) |
| $B_{eq}$ | 16d | -0.16(3) | - | - | 1.33(6) | - | - | 1.7(2) |
| | 16c | 0.6(1) | - | - | 1.4(1) | - | - | 1.2(2) |
| | $R_{wp}$ | 10.91 | - | - | 10.75 | - | - | 10.89 |
| | $\chi^2$ | 1.29 | - | - | 1.41 | - | - | 1.49 |

**Figure Captions**

Fig 1. Phase diagrams for each $Ln_2Ti_2O_7$ – $Ln_2TiO_5$ system (*Ln* = Gd-Lu), comparing known high temperature phases[38-42, 46-48] (upper rows) with those found in this study (lower rows). Pyrochlore (P) $Ln_2Ti_2O_7$ with either ordered hexagonal (HEX) or disordered fluorite (F) $Ln_2TiO_5$ two phase regions are seen for many of the larger lanthanides. Pyrochlore solid solutions, P (SSN), are formed at 1700 °C except for the Tb and Dy cases, where the series are quenched from the melt (> 1800 °C).

Fig 2. A simplified model of the original pyrochlore, $Ho_2Ti_2O_7$ (left), and the transition to a fluorite crystal structure (right) in the series $Ho_2(Ti_{2-x}Ho_x)O_{7-x/2}$. Ho (red) and Ti (blue) form separate corner sharing tetrahedral lattices in the pyrochlore structure, but develop edge sharing connectivity (green line) as extra Ho is stuffed in place of Ti. At x=0.67, Ho and Ti are completely mixed and can be represented in the fluorite structure as one site (purple) with disordered Ho/Ti in a 2:1 ratio. Oxygen atoms are omitted for clarity.

Fig 3. a) The lattice parameter, a, of the cubic pyrochlore materials, $Ln_2(Ti_{2-x}Ln_x)O_{7-x/2}$, with respect to stuffing level, *x* (main panel). Linear increase of the unit cell in each case reveals the effect of replacing $Ti^{4+}$ with larger $Ln^{3+}$ cations. Inset illustrates the slopes measured from the lattice parameter dependence on *x* (from main panel) plotted against the rare earth ionic radius. b) The lattice parameter, a, at three different *x* compositions plotted versus lanthanide ion radius. Dotted lines show a generally linear increase of unit cell size with varying rare earths.

Fig 4. Fraction of $Ti^{4+}$ occupancy on the pyrochlore *A*-site is plotted versus stuffing level. The small fraction of $Ti^{4+}$ mixing at low $x$ is taken to be zero within experimental error of our method. No mixing on the *A*-site shows that the stuffed $Ln^{3+}$ initially displaces $Ti^{4+}$ on the *B*-site before mixing more thoroughly on both sites at higher $x$ for *Ln* = Ho-Lu. Tb and Dy display minimal mixing on the *A*-site even at maximum stuffing.

Fig 5. Rietveld structure refinement of $Ho_2(Ti_{1.9}Ho_{0.1})O_{6.95}$ shown with raw data in blue, fitted pattern in red, and the difference in gray (main panel). A magnified view of the (2 2 2) peak for each of the stuffed Ho titanates is shown in the inset.

Fig 6. Rietveld structure refinement of $Dy_2(Ti_{1.9}Dy_{0.1})O_{6.95}$ shown with raw data in blue, fitted pattern in red, and the difference in gray (main panel). A magnified view of the (222) peak for each of the stuffed Dy titanates is shown in the inset. Peak broadening and asymmetry occur at higher levels of Dy stuffing.

Fig 7. Electron diffraction images of $Dy_2(Ti_{1.8}Dy_{0.2})O_{6.90}$ and $Dy_2(Ti_{1.33}Dy_{0.67})O_{6.33}$ show single phase materials with no evidence of short range *Ln*-Ti ordering in the structure. Lattice parameters measured here in the (400) direction are consistent with those determined from the XRD data.

Fig 1.

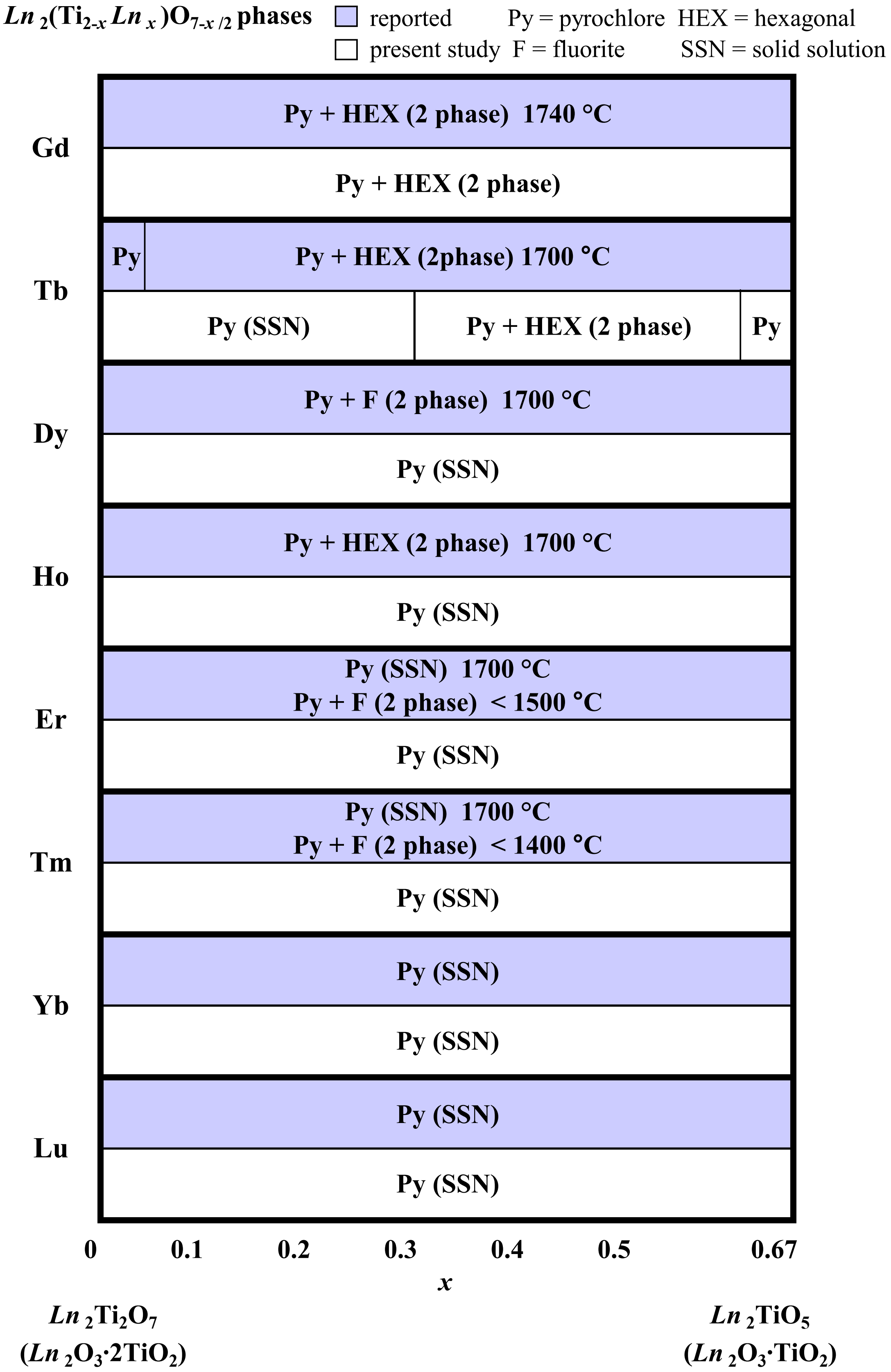

Ln 2(Ti2-x Ln x )O7-x /2 phases
reported
present study
Py = pyrochlore
HEX = hexagonal
F = fluorite
SSN = solid solution
Gd
Py + HEX (2 phase) 1740 °C
Py + HEX (2 phase)
Tb
Py
Py + HEX (2phase) 1700 °C
Py (SSN)
Py + HEX (2 phase)
Py
Dy
Py + F (2 phase) 1700 °C
Py (SSN)
Ho
Py + HEX (2 phase) 1700 °C
Py (SSN)
Er
Py (SSN) 1700 °C
Py + F (2 phase) < 1500 °C
Py (SSN)
Tm
Py (SSN) 1700 °C
Py + F (2 phase) < 1400 °C
Py (SSN)
Yb
Py (SSN)
Py (SSN)
Lu
Py (SSN)
Py (SSN)
0
0.1
0.2
0.3
0.4
0.5
0.67
x
Ln 2Ti2O7
(Ln 2O3·2TiO2)
Ln 2TiO5
(Ln 2O3·TiO2)

Fig. 2

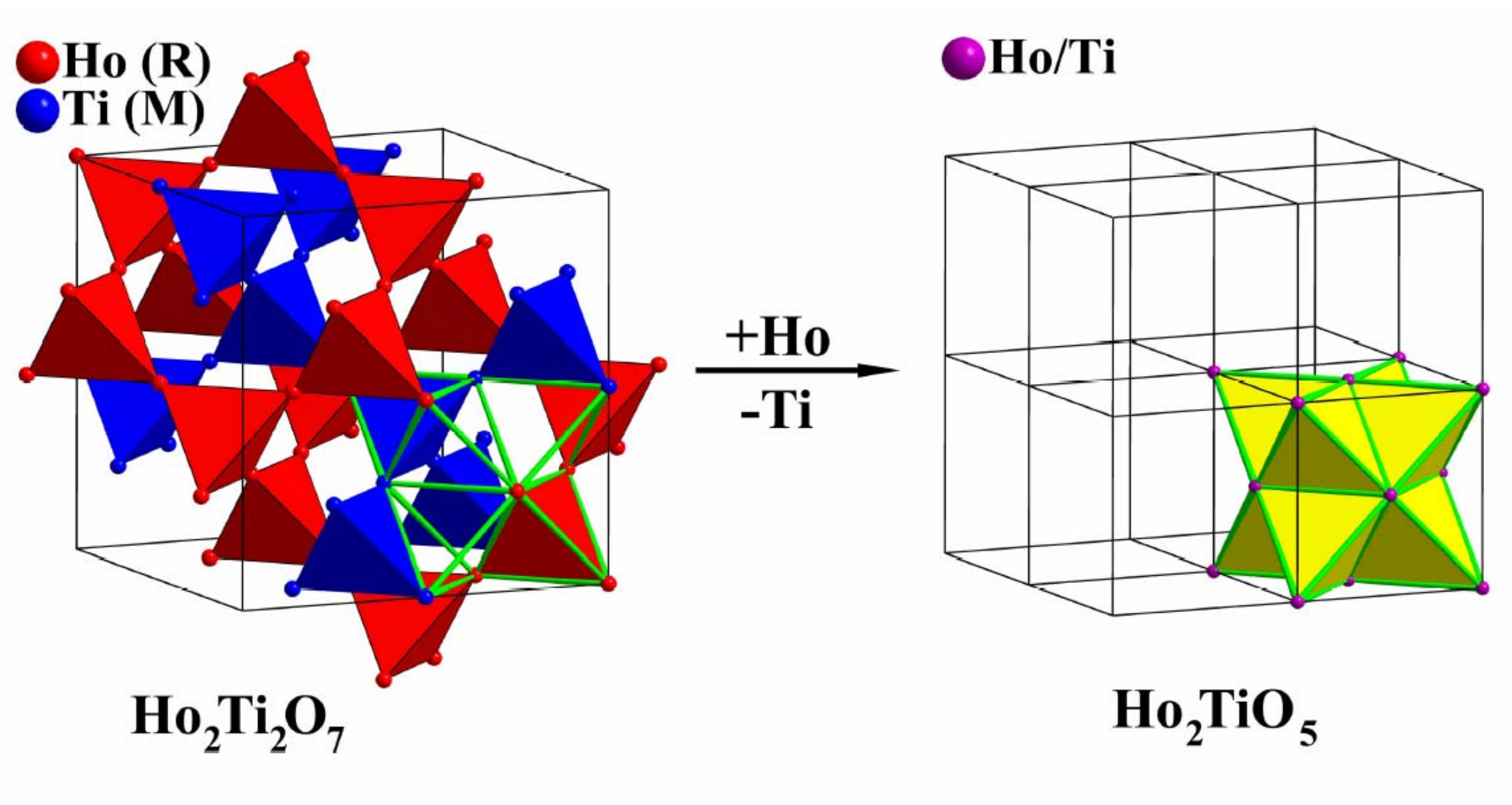

Fig. 3

a)

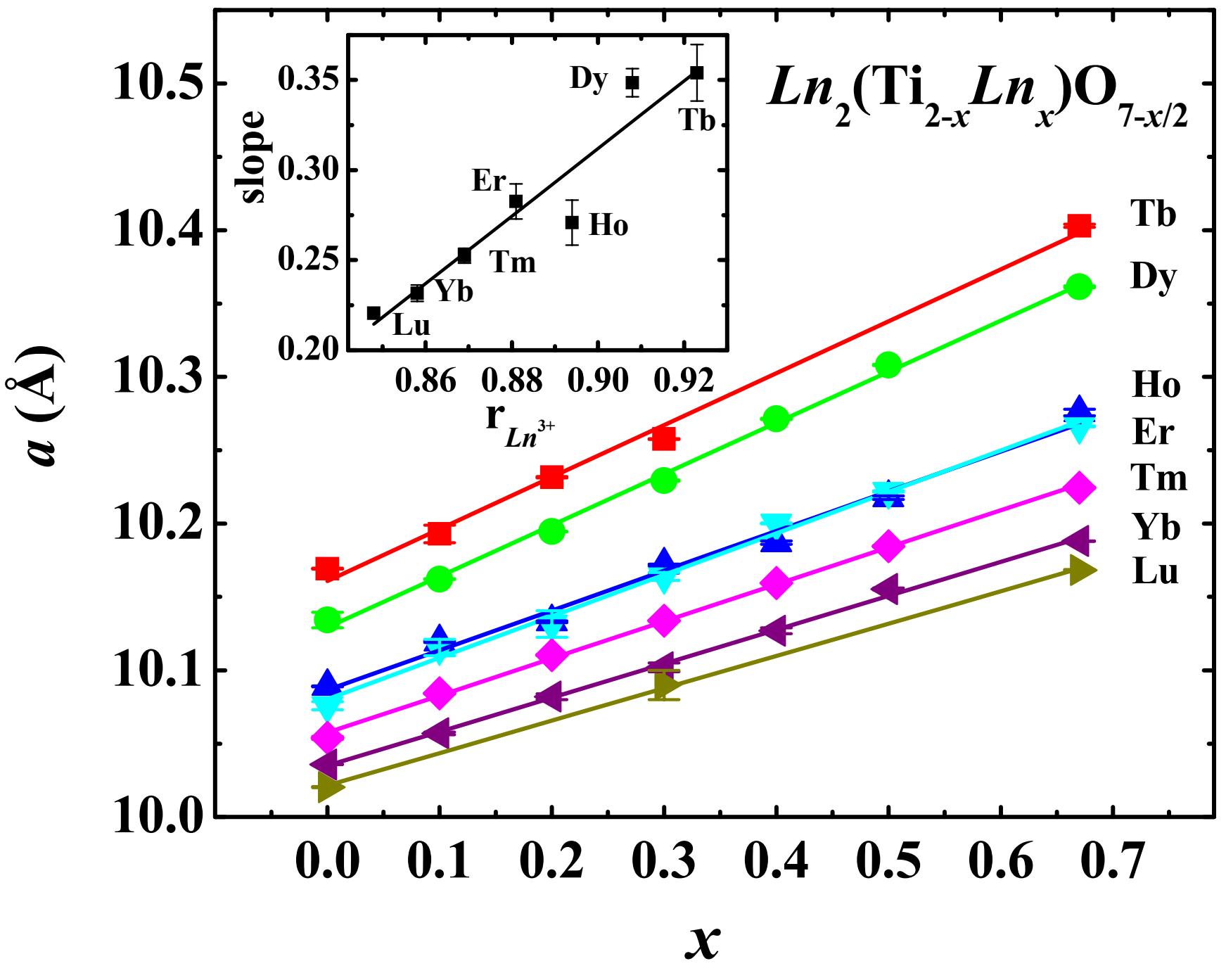

b)

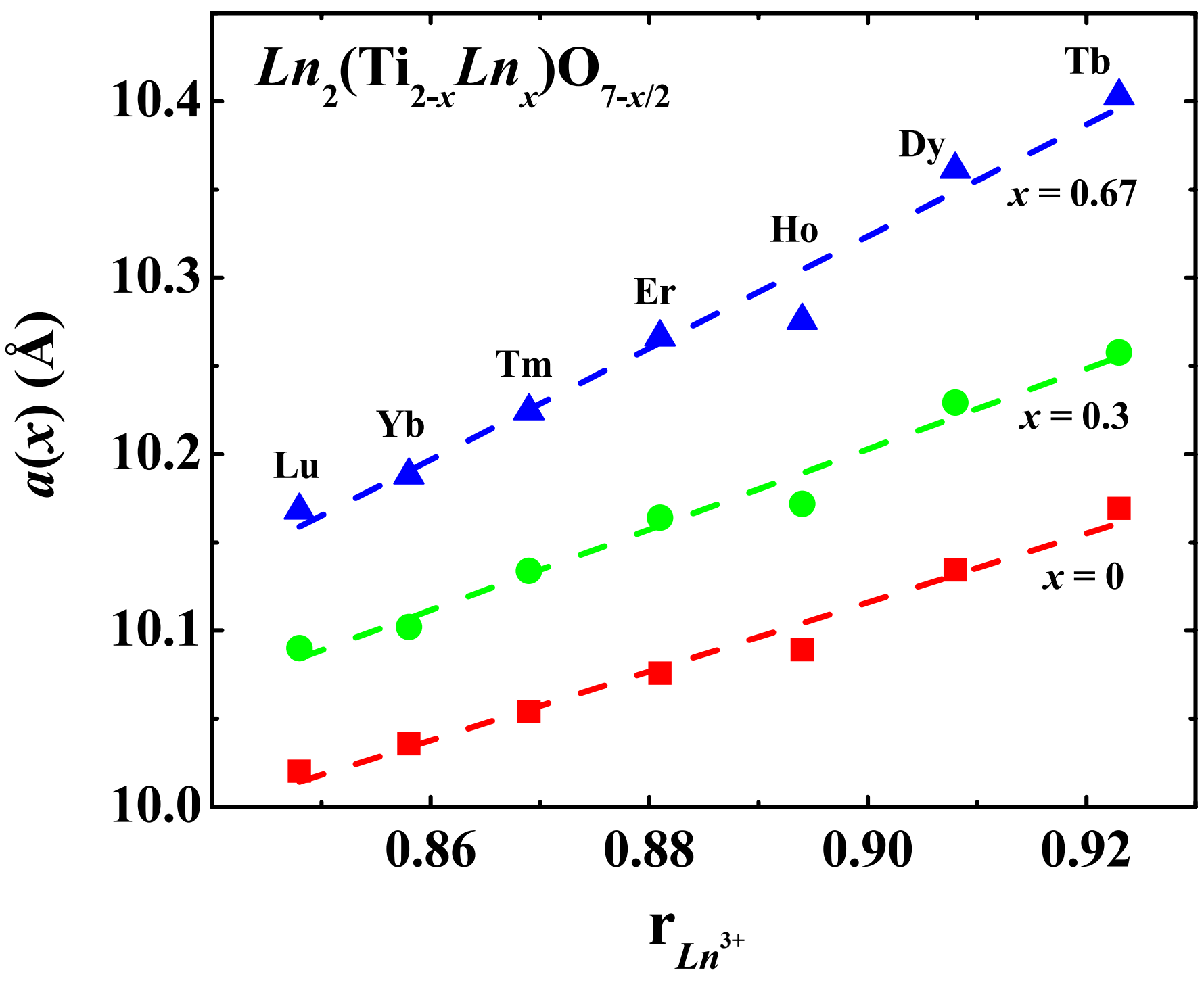

Fig. 4

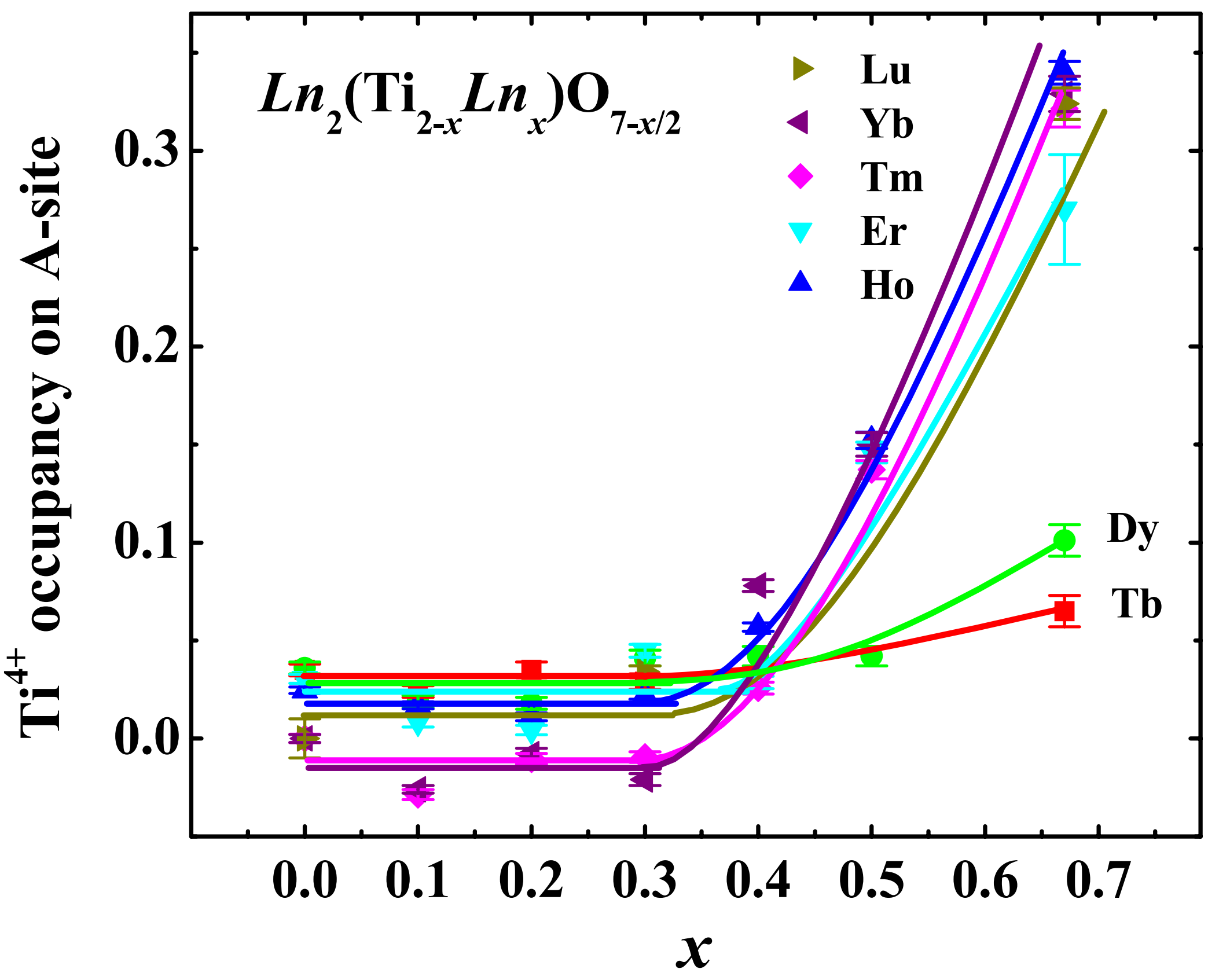

Fig. 5

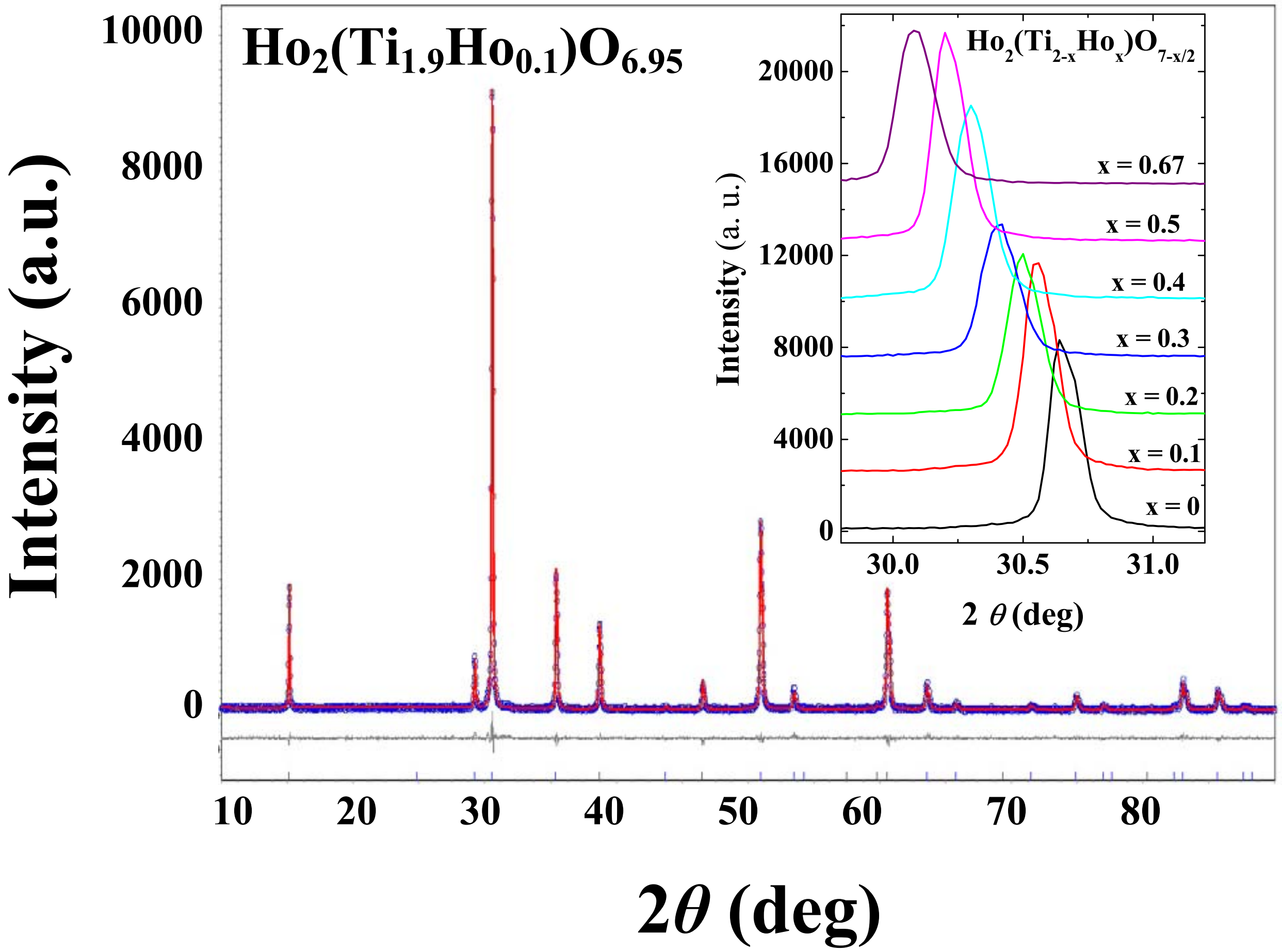

Fig. 6

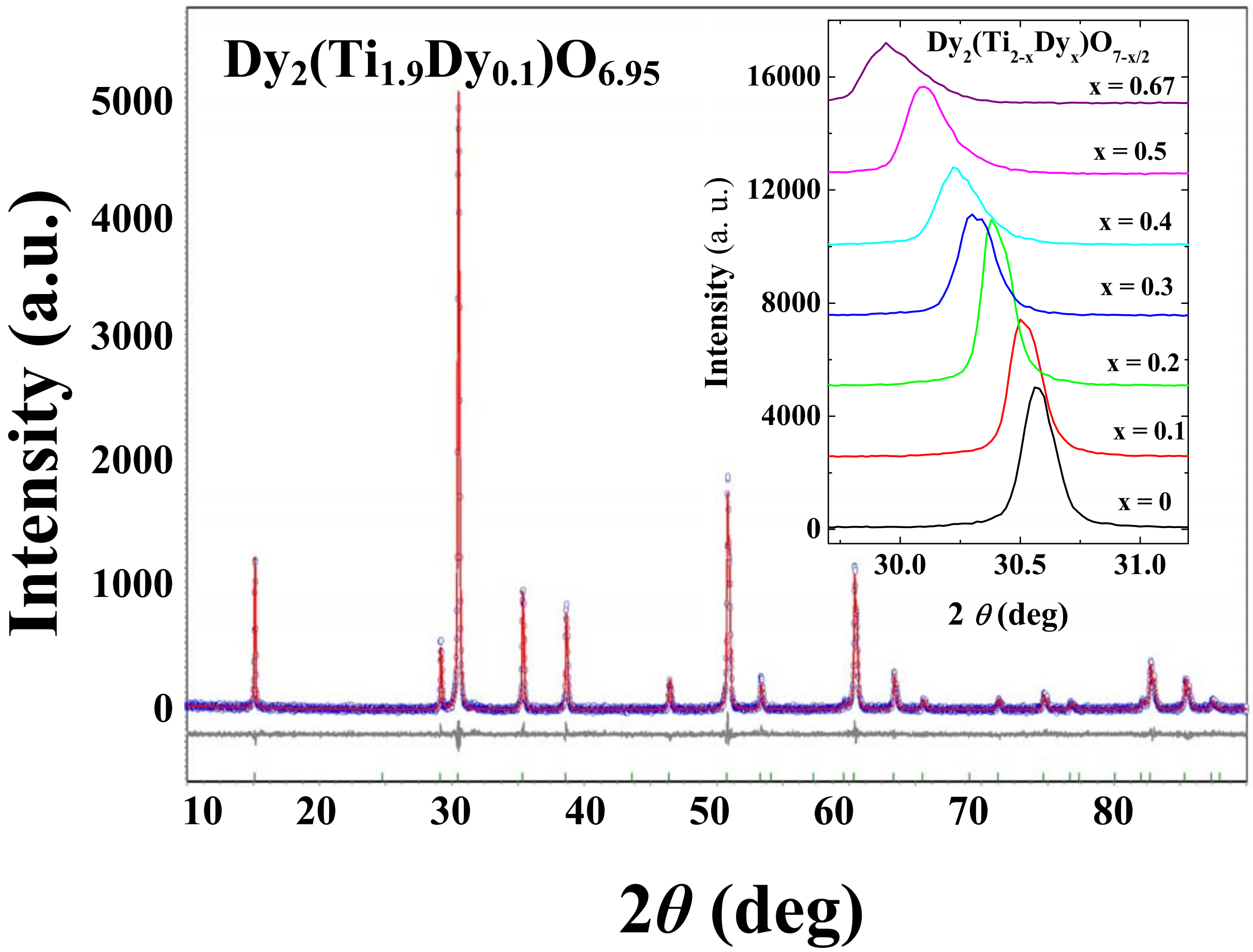

Fig. 7

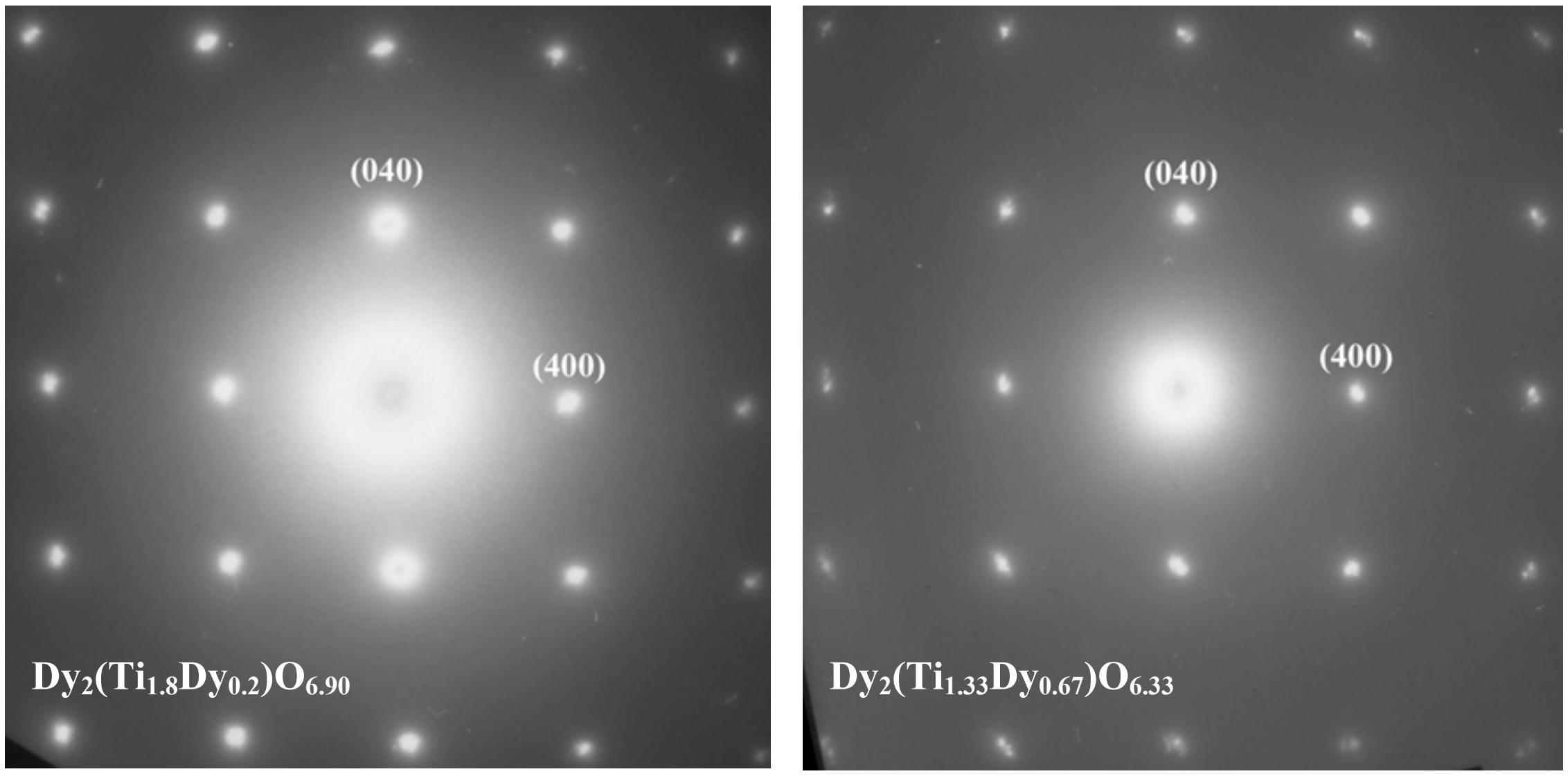